\documentclass{JHEP} 

\usepackage{epsfig}

\newcommand\fverb{\setbox\pippobox=\hbox\bgroup\verb}
\newcommand\fverbdo{\egroup\medskip\noindent%
			\fbox{\unhbox\pippobox}\ }
\newcommand\fverbit{\egroup\item[\fbox{\unhbox\pippobox}]}
\newbox\pippobox

\title{Hadron Production in Neutrino-Nucleon Interactions at High Energies}

\author{M. T. HUSSEIN, N. M. Hassan and W. Elharbi\\
Physics Department, Faculty of Science, Cairo University, 12613 Cairo, Egypt\\
E-mail: \email{husseinmt@hotmail.com}}

\received{\today} 		

\preprint{hep-ph/0105184}      

\abstract{
The multi-particle production at high energy neutrino- nucleon collisions are investigated
through the analysis of the data of the experiment CERN-WA-025 at neutrino energy less than 260GeV and 
the experiments FNAL-616 and FNAL-701 at energy range 120-250 GeV. The general features of these 
experiments are used as base to build a hypothetical model that views the reaction by a Feynman diagram of 
two vertices. The first of which concerns the weak interaction between the neutrino and the quark constituents 
of the nucleon. At the second vertex, a strong color field is assumed to play the role of particle production, 
which depend on the momentum transferred from the first vertex. The wave function of the nucleon quarks 
are determined using the variation method and relevant boundary conditions are applied to calculate the deep 
inelastic cross sections of the virtual diagram.} 


\begin{document} 

\maketitle 

\section{Introduction}

Observation of particles with large transverse momenta produced in
high-energy collisions of hadrons and nuclei provides information about the
properties of quark and gluon interactions. As follows from numerous studies
in relativistic physics [1, 2, 3], a common feature of the processes is the
local character of the hadron interactions. This leads to a conclusion about
dimensionless constituents participating in the collisions. On the other
hand, the deep inelastic interactions of leptons with nuclei may reveal
facts that relate the nature of particle production at lepton processes to
that produced by hadrons. All these reasons had persuaded us to pursue the
recent experiments of the neutrinos and antineutrinos with nucleons and
nuclei to extract the general features of such collisions that enable to set
up a relevant phenomenological model. The CERN experiment CERN-WA-025 [4],
and the Fermi National lab experiments FNAL-616 [5] and FNAL-701 [6] are
good samples to consider for examination. The first conclusion extracted
from these experiments is the logarithmic relation between the average
charged particle multiplicity and the lab energy of the incident neutrino as
demonstrated by Fig. (1). The figure shows also that neutrinos ($\nu $)
produce more particles than antineutrinos ($\overline{\nu }$) at the same
incident energy and in both cases the yield of particles is much less than
the corresponding case of hadron nucleon collisions [7]. Moreover, Fig. (2)
shows that the values of the 2nd moment does not depend on the type of the
projectile whether it is $\nu $ or $\overline{\nu }$ and this moment has the
exponential form similar to that of hadron nucleon. Finally, Figures (3 , 4)
show that F2 and x F3 are approximately independent on the momentum transfer
square Q2, but instead depend on the Bjorken scalar variable x.
In this work, a hypothetical model is proposed assuming a Feynman diagram of
two vertices. At the first of which, the incident neutrino interacts with 
one of the nucleon constituent quarks by weak interaction and exchanges a
heavy boson W or Z. The first vertex is considered as the source, which 
supply the second vertex with energy that is used to create pairs of hadrons
through strong interaction. Of course, the charged and neutral currents due
to W and Z in this case have their own property that is responsible to
different result of the hadrons. The paper is organized as follows: In section
2, we present the scenario of the model as well as the results and discussion.
Brief summary and conclusive remarks are given in section 3.

\section{The Proposed Model}

The problem of multi particle production in $\nu $-N collision is classified
into two main parts. The first is to propose a relevant diagram to describe
the process and specify the matrix element at its vertices. The second is to
specify the wave function of the constituent particles of the target. Our
strategy is to use a Feynman diagram as shown in Fig. (5) with two vertices.
At the first, a weak interaction takes place and a strong interaction holds
the second vertex. On the other hand, we follow the quantum variation method
[8] to determine the parton wave function. Consider a two-dimensional
Minkowski space, and we define the null momentum [9] $p$ as $p=p_{0}-p_{l}$
where $p_{o}$ represents the total energy and $p_{l}$ is the longitudinal
momentum, so that the mass shell condition for a particle of mass m: $%
p_{0}^{2}=p_{l}^{2}+m^{2}$ is replaced by $p_{0}=\frac{1}{2}(p+\frac{m^{2}}{p%
})$ . In QCD, the force between quarks is due to the exchange of gluons,
which are massless bosons, and corresponds to the photons in QED. The
long-range field between quarks may be considered as linear field in null
coordinates and has a propagator $1/q^{2}$ in momentum space. Let us
consider the nucleon as a relativistic quantum system of multiparticles. So
that the nucleon is described in terms of wave function that depends on
quantum numbers concerning the quarks (momentum, spin, color, flavor). The
values of these parameters are determined to minimize the total energy of
the quark system of the nucleon. Since the quarks forming the nucleon have
spin=1/2 and the nucleon is colorless, so it is convenient to write the
parton wave function as $\widetilde{\Psi }(a,\alpha ,p)$ , where $a$ is the
flavor (up or down), $\alpha $ is the color quantum number and $p$ is the
momentum in the null space. Then the nucleon wave function for $N$ partons
is $\widetilde{\Psi }(a_{1},\alpha _{1},p_{1};...;a_{N},\alpha _{N},p_{N})$
and should be anti-symmetric under interchange of a pair of quarks, as the
nucleons are Fermions. However, since the baryon is invariant under color,
the wave function must be completely in color alone

\begin{equation}
\widetilde{\Psi }(a_{1},\alpha _{1},p_{1};...;a_{N},\alpha _{N},p_{N})=\frac{%
\varepsilon _{\alpha _{1}...\alpha _{N}}}{\sqrt{N!}}\widetilde{\Psi }%
(a_{1},p_{1};...;a_{N},p_{N})
\end{equation}

The corresponding nucleon wave function in the position space is $\Psi (\{{%
a_{i},r_{i}\}})$ which is the Fourier transform of $\widetilde{\Psi }%
(a_{1},p_{1};...;a_{N},p_{N})$ . If $\mu _{a}$ is the quark mass and the
linear potential between a quark pair is $V(r_{i}-r_{j})=\frac{g^{2}}{2}%
(r_{i}-r_{j})$, where $g^{2}$ is the coupling constant, then the total
energy of the nucleon is

\begin{eqnarray}
\zeta _{N} &=&\sum_{a_{1}..a_{N}}\int_{0}^{\infty }\sum_{i=1}^{N}(p_{i}+%
\frac{\mu _{i}^{2}}{p_{i}})|\widetilde{\Psi }%
(a_{1}..a_{N};..p_{1}..p_{N})|^{2}\frac{dp_{1}..dp_{N}}{(2\pi )^{N}}+ 
\nonumber \\
&&+\frac{g^{2}}{2}\sum_{a_{1}..a_{N}}\int_{-\infty }^{\infty }\sum_{i\neq
j}^{N}(r_{i}-r_{j})|\Psi (a_{1}..a_{N};..r_{1}..r_{N})|^{2}dr_{1}...dr_{N}
\end{eqnarray}
Starting with the simplest parametric form of the quark wave function,

\begin{equation}
\widetilde{\Psi }(p)=C\exp (-ap)
\end{equation}
where $a>0$ and $p >= 0$ and the Fourier transform is

\begin{equation}
\Psi (r)=\frac{C^{\prime }}{(r+ia)^{2}}
\end{equation}

Where $C$and $C^{\prime }$are normalization constants, while $a$ is a fitting
parameter. It is found that $ a= 0.5 $ gives minimum value of the total energy 
as shown in Fig.(6). 

A more reasonable quark wave function with two variation parameters $\alpha $%
and $\beta $, is

\begin{equation}
\widetilde{\Psi }(p)=Cp^{\alpha }(1-p)^{\beta }
\end{equation}

Again, inserting Eq. (2.4) in Eq. (2.1), this results that the total energy $%
E_{tot}$of the quark assembly of the nucleon shows multiple minima
corresponding to different energy levels whose eigen parameters $\alpha $and 
$\beta $are listed in Table (1). The range of the null momentum $p$extends
from zero up to $P_{max}$. It is more convenient to express the wave
function and all other physical quantities in terms of the scaling Bjorken
variable $x=p/P_{\max }$where $0<x<1$.
\begin{center}
Table (1)

\begin{tabular}{|l|l|l|l|}
\hline
Parton state & $\alpha $ & $\beta $ & Energy (arbitrary unit) \\ \hline
E0 & 3.6 & 3.6 & 3.45x10$^{-6}$ \\ \hline
E1 & 2.9 & 3.7 & 6.76x10$^{-6}$ \\ \hline
E2 & 2.7 & 3.7 & 9.54x10$^{-6}$ \\ \hline
E3 & 2.2 & 3.4 & 4.60x10$^{-6}$ \\ \hline
E4 & 2.1 & 3.1 & 4.70x10$^{-6}$ \\ \hline
\end{tabular}
\end{center}

The ground state E0 has equal values in the parameters $\alpha $and $\beta $%
which means that the state has line symmetry about $x=0.5$, in other words,
the probability that a parton has a value $x$is equal to that of $(1-x)$.
The symmetry breaks at the higher-up states, where the probability increases
towards the deep inelastic $x<0.5$. The wave functions defined by Eqs. (2.2)
and (2.4) are to be used to calculate the scattering amplitude $\Gamma $due
to the diagram Fig. (5) as,

\begin{equation}
\Gamma _{tot}=\prod_{i}\Gamma _{i}
\end{equation}

where $\Gamma _{i}$is the scattering amplitude at the i$^{th}$vertex of the
diagram. Since the diagram contains only two vertices, where a weak
interaction holds the first and a strong one at the second, and each one is
associated with its relevant propagator, so that $\Gamma _{tot}$may be
written as,

\begin{equation}
\Gamma _{tot}(x,q^{2})=\frac{1}{q^{2}+M_{W,Z}^{2}}\exp (-\gamma q^{2})|%
\widetilde{\Psi }(x)|^{2}
\end{equation}

Eq. (2.6) includes three factors, the first is the propagator of the weak
field in which $M_{W,Z}$represents the mass of the exchange Boson [10] for
the charged $W^{\pm }$or the neutral $Z^{0}$current. The second factor is
the propagator for the strong field or sometimes known as the profile
function [11]. The third factor is the probability density. The problem of
particle production is to be viewed through the relativistic phase space.
Following the scheme of Ref. [12], it is easy to define the cross section of
finding n-particles in the final state with total center of mass energy $W$, 
$W^{2}=(p_{a}+p_{b})^{2}$

\begin{equation}
\sigma _{n}=\frac{1}{F}\int ...\int \prod_{i}\{\frac{d^{3}p_{i}}{2E_{i}}%
\}\delta ^{4}(p_{a}+p_{b}-\sum p_{i})|\Gamma (p_{i})|^{2}
\end{equation}
Where $F$represents the incident flux and the delta function is to conserve
the four-vector momentum at the $2^{nd}$vertex of the reaction. It also
restricts the integration over a surface of 3n-4dimensional space. The
multi-dimensional integration in Eq. (2.7) may be solved using the Monte
Carlo technique. However the volume of such a sphere may be approximated at
extremely high energy as,

\begin{equation}
\sigma _{n}=\frac{1}{F}\frac{(\pi /2)^{n-1}}{(n-1)!(n-2)!}W^{n-2}|\Gamma
(p)|^{2}
\end{equation}

Transforming Eq. (2.7) from the momentum space to the Bjorken scaling $x$and
writing the energy $W$in terms of the kinematical variables of diagram (5), $%
W^{2}=m^{2}+q^{2}(\frac{1}{x}-1)$, then

\begin{equation}
\sigma _{n}(x,q^{2})=\frac{1}{F}\frac{(\pi /2)^{n-1}}{(n-1)!(n-2)!}[%
m^{2}+q^{2}(\frac{1}{x}-1)]^{n-2}|\Gamma (x,q^{2})|^{2}
\end{equation}

The implementation of Eq. (2.9) results in Fig. (7), which shows that the
cross section of production of n-particles depends appreciably on $x$and
weakly on $q^{2}$. The decrease of $x$means going towards the deep
inelastic, which offer more fraction of energy for creation of particles and
this increases the peak of the multiplicity towards higher values. The
overall multiplicity distribution is found by integrating over x from 0 to 1
and over $q^{2}$from . A cutoff value of $q^{2}$is necessary to keep the
system in the physical region of the phase space. The cutoff ratio is taken
as a free parameter in this model to fit the experimental data. The
multiplicity distributions of secondary hadrons are demonstrated in Fig. (8)
for ?-energies of 2, 8, 22 and 146 GeV, the average values of each are
listed in Table (2) with the considered parameters. The comparison with the
experimental data shows very good agreement. The cross section $\sigma
_{n}(x,q^{2})$is recalculated again using the parton wave function as in Eq.
(2.4) for the first 3- parton states only E0, E1 and E2 with the
corresponding parameters as in Table (1). The results are given in Table (3)
for the average multiplicity compared with the experimental values that show
also good agreement. In fact the comparison with the average value of the
multiplicity is not enough to reflect the validity of a hypothetical model,
nevertheless, it may indicate that the work is going in the right way. In
the forth- coming paper we managed to take into account the effect of the
exchange whether it is the neutral boson $Z^{0}$or the charged $W^{\pm }$.
This may be used to interpret the increase of the yield in $\nu $-N
collisions than the case of $\overline{\nu }$-N. Also the calculation of the
differential cross section $d^{2}\sigma /dxdq^{2}$may give more information
on such reactions.
\begin{center}
Table (2)  \\
\begin{tabular}{|c|c|c|c|c|}
\hline
Cutoff & E$_{\nu }$ & a & {<}n{>}$_{\exp }$ & 
{<}n{>}$_{th}$ \\ \hline
0.075 & 2 & .5 & 1.97 & 2.1 \\ \hline
0.070 & 8 & .5 & 3.30 & 3.31 \\ \hline
0.020 & 22 & .5 & 4.50 & 4.35 \\ \hline
0.001 & 146 & .5 & 6.14 & 6.17 \\ \hline
\end{tabular}

\smallskip 

\smallskip 
Table (3)\\
\begin{tabular}{|c|c|c|c|c|}
\hline
{<}n{>}exp  & $\gamma $(E0) & $\gamma $(E1) & $\gamma $%
(E2) & {<}n{>}th \\ \hline
1.97 & 1 & 1.1 & 1.15 & 2.10 \\ \hline
3.30 & 0.31 & 0.36 & 0.37 & 3.31 \\ \hline
4.50 & 0.20 & 0.26 & 0.27 & 4.35 \\ \hline
6.14 & 0.03 & 0.033 & 0.033 & 6.17 \\ \hline
\end{tabular}
\end{center}

\section{Concluding Remarks.}

\begin{itemize}
\item  The average multiplicity of hadrons produced in $\nu $-nucleon
interactions has a logarithmic dependence on the energy of the incident
neutrino. The yield in $\nu $-N is slightly greater than the corresponding
figure in $\overline{\nu }$-N.

\item  The $\nu $-N interaction may successfully described by a Feynman
diagram of two vertices. A weak interaction holds the first and a strong
interaction at the second where hadrons are created. W and Z bosons are
assumed as the exchange particles in the weak interaction.

\item  The first vertex of the Feynman diagram is assumed as the source,
which supply the energy to the second vertex. That energy is responsible for
hadron creation. The rate of energy transfer is controlled by the matrix
element of weak interaction at the first vertex.

\item  The average multiplicity of produced hadrons is strongly dependent on
the Bjorken scaling variable $x$ and weakly on the momentum transfer $q^{2}$.
\end{itemize}

\EPSFIGURE{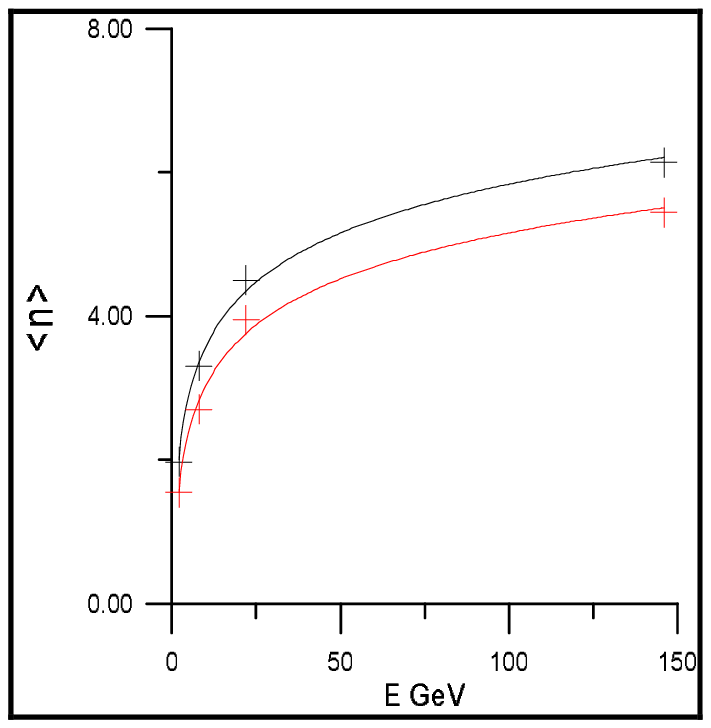,width=.6\textwidth}{Average multiplicity of hadrons
produced in $\nu-D$ (black +) and $\overline{\nu }-D$ (red +). The data concern
the experiment CERN-WA-025. The solid lines are the parametric fitting curve
of the form of $<n> = 0.981 * log(E) + 1.318 $ for $\nu-D$ experiment and 
$<n> = 0.922 * log(E) + 0.909$ for the $\overline{\nu }-D$.}

\EPSFIGURE{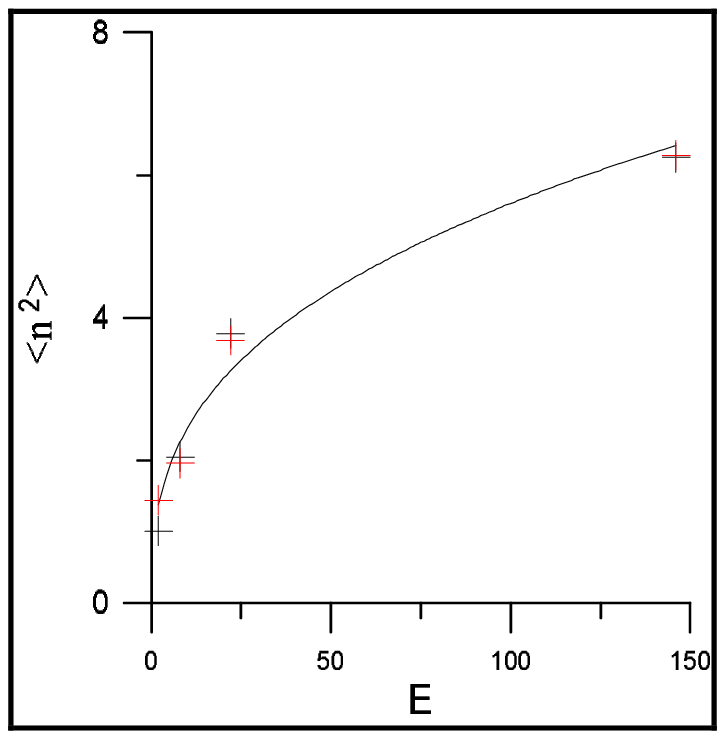,width=.6\textwidth}{
The second moment of the multiplicity
of hadrons produced in $\nu-D$ (black +) and $\overline{\nu }-D$ (red +). 
The data concern the experiment CERN-WA-025. 
The solid line is the parametric fitting curve of the 
form $<n2> = 1.075* E^{.358}$.} 

\EPSFIGURE{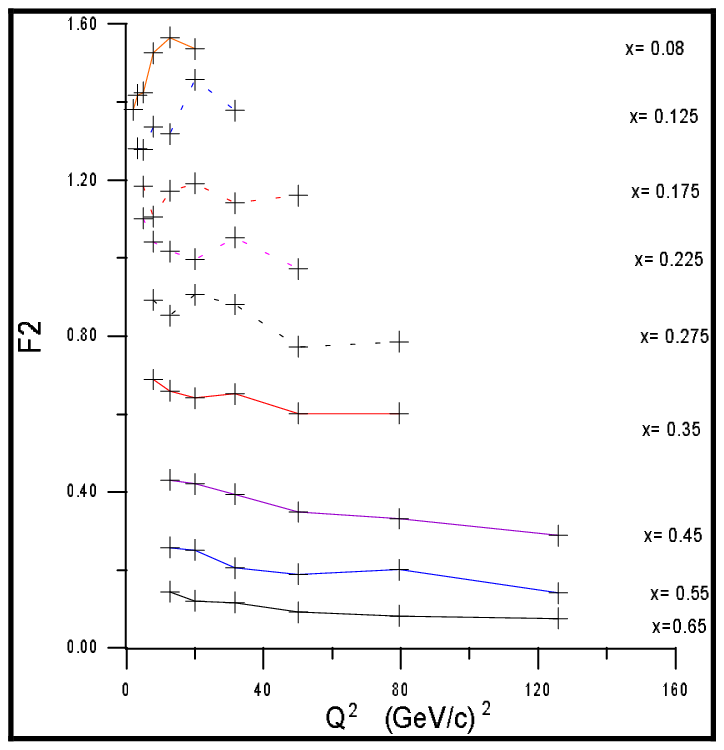,width=.6\textwidth}{
The structure function F2 of the ($\nu p$)
collision of the experiments FNAL-616 and 
FNAL-701 at energy range 120-250 GeV, with 
Bjorken scaling variable $0.08 < x <0.65$.}

\EPSFIGURE{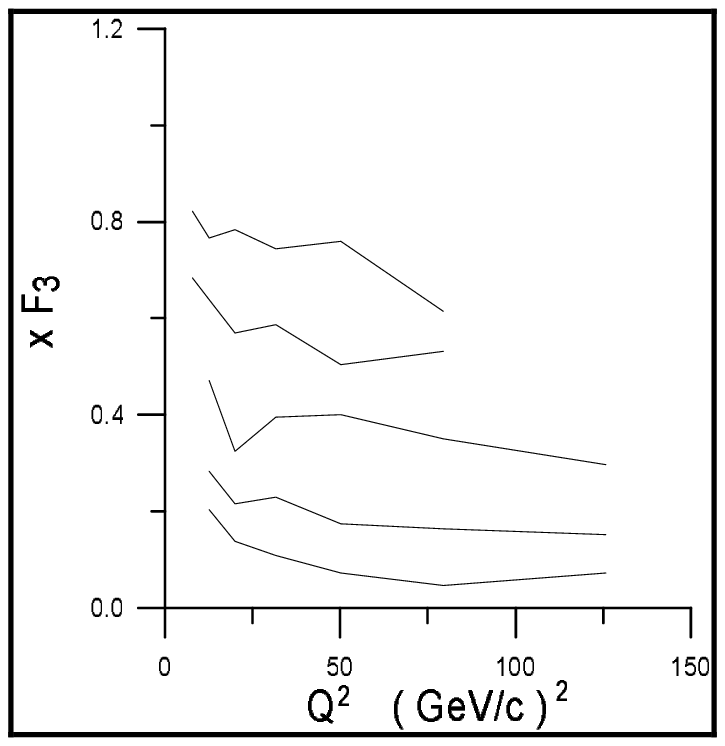,width=.6\textwidth}{
The structure function xF3 of the ($\nu p$)
collision of the experiments FNAL-616 and 
FNAL-701 at energy range 120-250 GeV, with 
Bjorken scaling variable $0.08 < x <0.65$.}

\EPSFIGURE{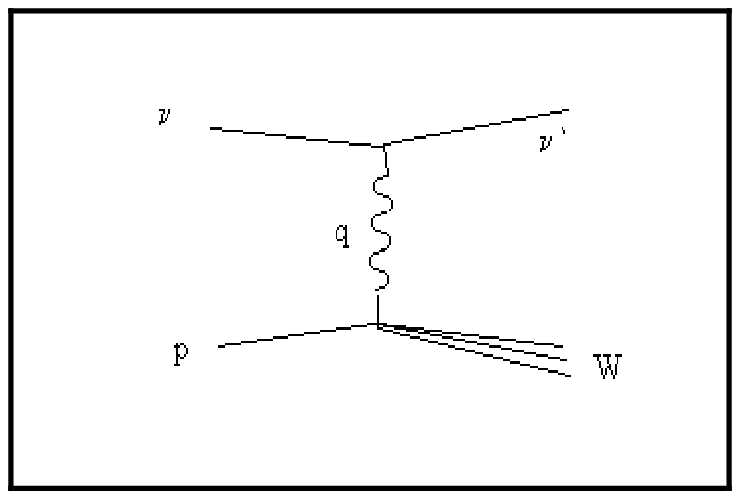,width=.6\textwidth}{
Feynman diagram describes the
multiparticle production in $\nu-N$ collisions.}  

\EPSFIGURE{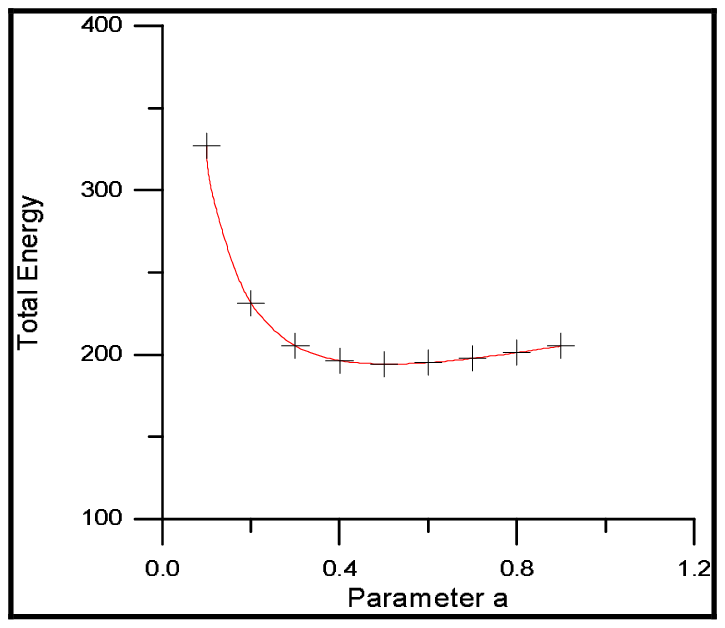,width=.6\textwidth}{
The variation of the total energy of the
assembly of the quarks forming the nucleon shows 
that it has a minimum at a parameter value $a = 0.5$}

\EPSFIGURE{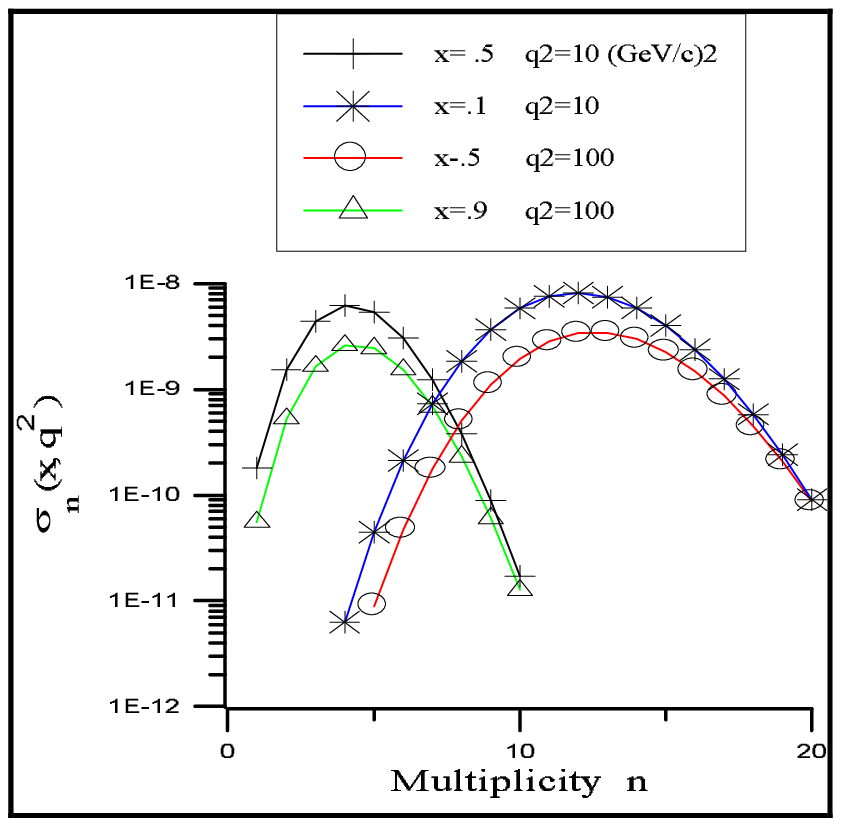,width=.6\textwidth}{
The multiplicity distribution of charged
secondaries in $\nu p$ collision as a function of both x 
and $q^{2}$.}

\EPSFIGURE{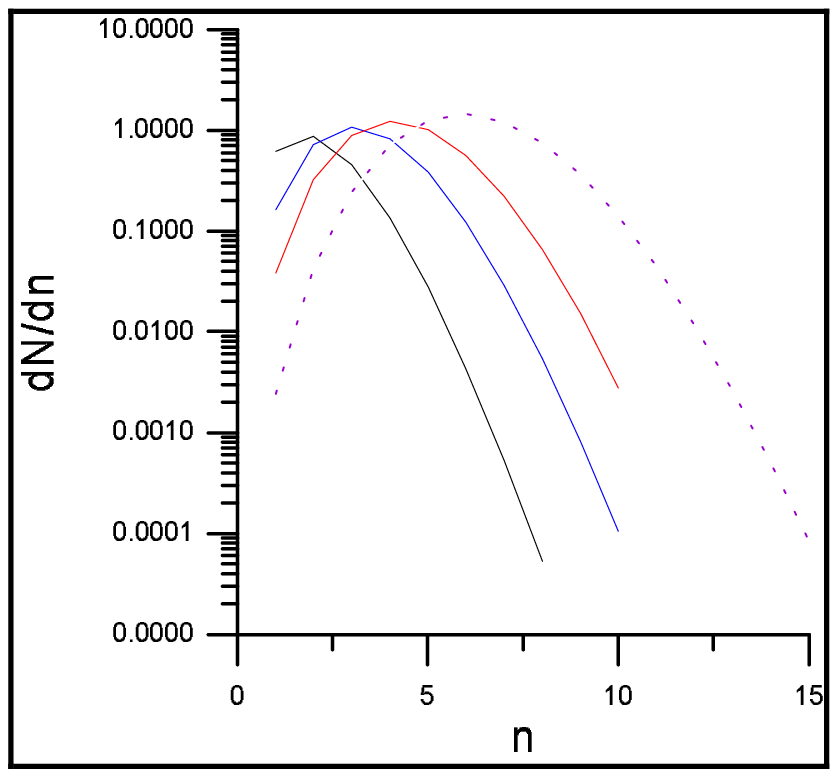,width=.6\textwidth}{
The overall multiplicity distribution of
charged secondaries produced in $\nu p$ collision as 
predicted by the model.}

\end{document}